\documentclass[useAMS,usenatbib]{mn2e}

\usepackage[dvips]{graphics}
\def \mdot   {{\hbox{$\skew3\dot M$}}}

\title[Clumped fast wind of NGC6543]{Structure and clumping in the fast 
wind of 
NGC6543}
\author[R.K. Prinja et al.]{R. K. Prinja$^{1}$\thanks{E-mail: 
rkp@star.ucl.ac.uk (RKP); seh@star.ucl.ac.uk (SEH); 
massa@taotaomona.gsfc.nasa.gov (DLM); fullerton@stsci.edu (AWF);
awxb@star.ucl.ac.uk (AWB)},
S. E. Hodges$^{1}$,
D. L. Massa$^{2}$,
A. W. Fullerton$^{3}$, and
A. W. Burnley$^{1}$\\
$^{1}$Dept. of Physics {\&} Astronomy, University College London, Gower Street, London WC1E 6BT \\
$^{2}$SGT, Inc., NASA Goddard Space Flight Center, Code 681.0, Greenbelt, MD 20771 \\
$^{3}$Space Telescope Science Institute, 3700 San Martin Drive, Baltimore, MD 
21218
}
\begin{document}

\date{Accepted 2007. Received 2007; in original form 2007}

\pagerange{\pageref{firstpage}--\pageref{lastpage}} \pubyear{2007}

\maketitle

\label{firstpage}

\begin{abstract}
Far-UV spectroscopy from the $FUSE$ satellite is analysed to {\it uniquely}
probe spatial structure and clumping in the fast wind of the
central star of the H-rich planetary nebula NGC~6543 (HD~164963).
Time-series data of the unsaturated P{\sc v} $\lambda\lambda$1118, 1128
resonance line P~Cygni profiles provide a very sensitive diagnostic
of variable wind conditions in the outflow. We report on the discovery
of episodic and recurrent optical depth enhancements in the P{\sc v}
absorption troughs, with some evidence for a $\sim$ 0.17-day modulation
time-scale. SEI line-synthesis modelling is used to derive
physical properties, including the optical depth evolution of
individual `events'. The characteristics of these features are
essentially identical to the `discrete absorption components' (DACs) 
commonly seen in the UV lines of massive OB stars. We have also employed
the unified model atmosphere code CMFGEN to explore spectroscopic
signatures of clumping, and report in particular on the
clear sensitivity of the P{\sc v} lines to the clump volume filling
factor. The results presented here have implications for the
downward revision of mass-loss rates in PN central stars. We conclude
that the temporal structures seen in the P{\sc v} lines of NGC~6543 likely
have a physical origin that is similar to that operating in
massive, luminous stars, and may be related to 
near-surface perturbations caused by stellar pulsation and/or magnetic 
fields.
\end{abstract}


\section{Introduction}

As the immediate precursors of white dwarfs, the central stars of
planetary nebulae (CSPNs) provide fundamental tests in key areas of
stellar astrophysics, including evolutionary theory, non-LTE model
atmospheres and plasma hydrodynamics. An important challenge is
to understand details of the mass-loss process through the planetary
nebula stage. Specifically, the fast winds of CSPN provide a probe
of the {\it current} mass-loss, and represent a valuable setting for the
study of radiative and mechanical interactions between stars and their
environments.

Shortly after the launch of the International Ultraviolet Explorer
($IUE$) satellite, it became clear that fast winds are a common
feature of CSPN (e.g. Heap, 1979;
Perinotto, Cerruti-Sola, Benvenuti, 1982).
The P~Cygni UV
resonance line profiles have been decoded to record maximum wind 
velocities of up to a few 1000 km s$^{-1}$ and estimates of the
mass-loss rate range between $\sim$ 10$^{-11}$ to 10$^{-6}$
M$_\odot$ yr$^{-1}$ (see e.g. Perinotto, 1989 and references within).
Our objective in this paper is to exploit new Far-Ultraviolet
Spectroscopic Explorer ($FUSE$) satellite time-series spectroscopy
of the central star of NGC~6543 to uniquely examine the
time-variable characteristics of its fast wind.

An improved understanding of variability and clumping in the
fast winds of CSPN is important for several reasons:
Temporally variable spatial structure implies a clumped and
at least locally non-spherical fast wind, which can modify how the
supersonic outflow interacts with the nebular material (e.g.
Falle et al. 2002; Steffen {\&} Lopez 2003). Substantial wind
clumping can also significantly complicate determinations of the
mass-loss rates from the central stars. 
For example, recent analyses of optical
and UV lines in luminous, massive O-type stars have provided evidence
for extreme wind clumping factors, with implications that the mass-loss
rates of OB stars may have previously been over-estimated by factors
of between 3 to 10 (e.g. Hillier, Lanz {\&} Heap 2003; Massa et al. 2003; 
Markova et al. 2004; Bouret, Lanz {\&} Hillier 2005; Fullerton, Massa 
{\&} Prinja 2006).

Recent $Chandra$ and $XMM$$-$$Newton$ observations have established the
presence of diffuse X-ray emission from {\it within} the nebular
interior regions (e.g. Guerrero et al. 2001; Kastner et al. 2003;
Maness {\&} Vrtilek, 2003). The source of the X-rays from the
central star vicinity remains uncertain, and one possible origin would
be the presence of shock-heated gas arising from instabilities in
a variable fast wind (e.g. Akashi, Soker {\&} Behar, 2006).

The present study is also motivated by a need to understand the
relation between optical spectroscopic (radial velocity)
and photometric changes evident in several CSPN. The frequently
imaged non-spherical and bipolar morphologies of the planetary
nebulae (e.g. Zuckerman {\&} Aller, 1986) raise the possibility that
the planetary nebulae may have been ejected via common envelope
interaction in close binary central stars (e.g. Sandquist et al.
1998). Surveys of the PN binary population are currently frustrated
by the physical interpretation of the observed radial velocity
and photometric variations, which might at least in part be due to
changes in the fast wind (e.g. Handler, 2003; De Marco et al. 2004).


\begin{table}
 \centering
\caption{NGC~6543 central star parameters.}
  \begin{tabular}{lll}
  \hline
Parameter & Value & Reference  \\
\hline

Sp. type & Of-WR (H-rich) & Mendez et al. (1990) \\
Luminosity & 5200 L$\odot$ & de Koter et al. (1996) \\
T$_{\rm eff}$ & 63000 K  & Georgiev et al. (2006) \\
Radius & 0.6 R$\odot$ & Georgiev et al. (2006) \\
Distance & 1001 $\pm$ 269 pc & Reed et al. (1999) \\
Mass-loss rate & $\sim$ 1$\times$10$^{-7}$ M$\odot$/yr & de 
Koter et al. (1996); \\
 & & Georgiev et al. (2006) \\
Terminal velocity & 1400 km s$^{-1}$ & This study. \\
Wind flushing time & 45 mins & This study. \\
\hline

\end{tabular}
\end{table}


There is a clear need therefore to investigate further the detailed
temporal variability of the fast winds in CSPN. To date, studies of 
variability in these outflows have primarily relied on (limited) multiple
UV spectra obtained with the $IUE$ satellite (e.g. Patriarchi {\&}
Perinotto, 1997). Generally however the detection of wind line
variability with $IUE$ (or $HST$) was extremely difficult since the only
UV resonance lines accessible are strongly saturated, thus masking
all changes except at the extreme violet edges of the line profiles.
In addition, the exposure times of high resolution $IUE$ spectra
were rather long, i.e. $\sim$ 2 to 3 hours, which is in fact
comparable to the wind flushing time in these stars.
We provide here new
perspectives on variability in the fast winds of CSPN by exploiting
$FUSE$ ($\lambda\lambda$905 to 1187 {\AA}) time-series observations.
$FUSE$ can uniquely deliver the requisite high signal-to-noise data
in short integration times, while also potentially providing access to
{\it unsaturated} resonance lines, including P{\sc v},
S{\sc iv}, S{\sc vi}, O{\sc vi}, and C{\sc iii}.

\begin{figure}
\centering
 {\scalebox{0.57}{\includegraphics{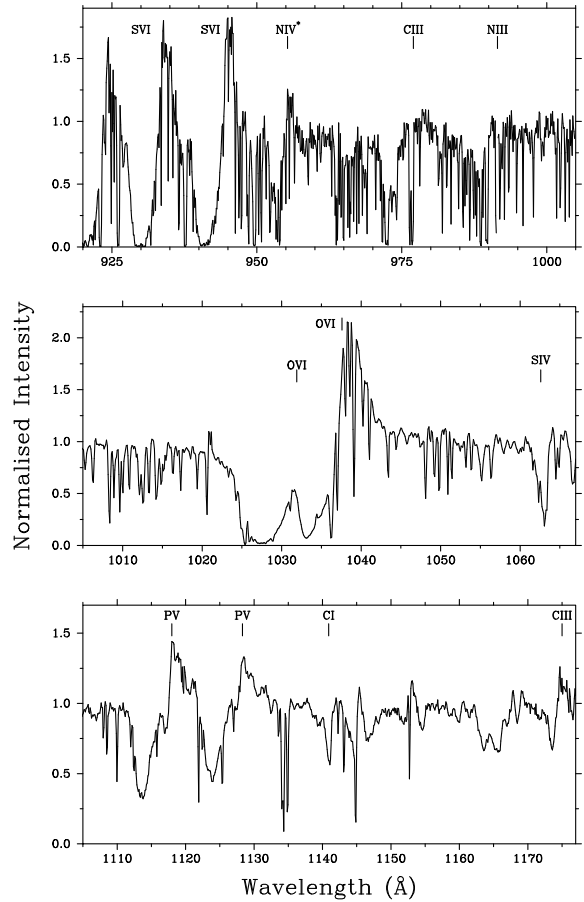}}}
 \caption{Mean FUSE spectrum of the central star in NGC~6543 from the F034 
run in January 2007.}
\end{figure}

\subsection{The target, NGC~6543}

We present in this paper a study of structure and variability in
the fast wind of the central star of the planetary nebula NGC~6543
(HD~164963; BD +66$^o$ 1066). The nebula is well studied,
with $HST$ images in particular revealing complex structures that
include phenomena such as bubbles and precessing jets
(e.g. Harrington {\&} Borkowski 1994; Balick 2004;
Wesson {\&} Liu, 2004).
X-ray emission has been detected with $Chandra$ (Chu et al. 2001;
Guerrero et al. 2001), which is consistent with a point source
at the central star and diffuse emission from within the
elliptical central shell. Various studies have reported evidence
for time-variable changes associated with the central star, including
Mendez, Herrero {\&} Manchado (1990; optical spectroscopy),
Bell, Pollacco {\&} Hilditch (1994; photometry), and
Patriarchi {\&} Perinotto (1997; UV spectroscopy).
Fundamental parameters of the central star based on non-LTE
model atmosphere analysis have been derived by
de Koter et al. (1996; ISA-Wind code) and Georgiev et al.
(2006; CMFGEN). Key adopted parameters for the central star in
NGC~6543 are listed in Table 1.
We note that for a '$\beta$=1'-type wind velocity law (e.g. Sect. 5),
the wind flushing time to travel from 0.1 to 0.9 of the
terminal velocity ($v_\infty$) is only $\sim$ 45 mins.

\section{FUSE time-series data}

Our study of the fast wind of NGC~6543 is largely based on $FUSE$
data from Program F034 (P.I. $-$ D.L. Massa), secured between 2007 
January 13 to 16.
A summary log of observations is provided in Table 2.
The observations were obtained through the MDRS (4$^"$ $\times$
20$^"$) aperture, spanning a total wavelength range of
905 to 1187 {\AA} at a spectral resolution of $\sim$ 15 km s$^{-1}$.
(The instrument and performance is discussed by Moos et al. 2000
and Sahnow et al. 2000.)
Typical individual integration times in HIST mode are $\sim$ 10 minutes.
Using the MDRS aperture unfortunately meant that occasionally
the target was not fully captured and there were cases of light loss.
The most usable components are 59 spectra (processed through
CalFUSE version 3.1.8) in the LiF2 channel (segment A; 
$\lambda\lambda$1086 to 1182 {\AA})
which covers the strategic line of
P{\sc v} $\lambda\lambda$1117.98, 1128.01.
A sparser sequence of 29 spectra for the LiF1 channel
(segment A; $\lambda\lambda$987 to 1082 {\AA}) provided some information
on the behaviour of the (stronger)
O{\sc vi} $\lambda\lambda$1031.92, 1037.62 wind-formed line.


\begin{table}
 \centering
\caption{Summary log of $FUSE$ observations.}
  \begin{tabular}{lllll}
  \hline
Obs. ID & No. of exp. & UT Date & MJD(start) & $\Delta$T (hours) 
\\
\hline
F0340105 & 26 & 2007-01-13 & 54113.1260 & 9.7 \\
F0340106 & 18 & 2007-01-14 & 54114.1097 & 5.9 \\
F0340107 & 19 & 2007-01-15 & 54115.0933 & 7.3 \\
F0340108 & 9 & 2007-01-16 & 54116.0680 & 1.4 \\
\hline

\end{tabular}
\end{table}


The mean $FUSE$ spectrum is shown in Figure 1. Aside from P{\sc v}
and O{\sc vi}, a strong P~Cygni line is also seen in
S{\sc vi} $\lambda\lambda$933.38, 944.52. There is some evidence
for blue-ward absorption in the excited N{\sc iv} $\lambda$955.34
line profile
and C{\sc iii} $\lambda$1175.67. The remainder of the spectrum is 
dominated
by narrow interstellar and circumsystem absorption lines
due to atomic species and molecular hydrogen.

Overall the mean $FUSE$ spectrum of the central star in NGC~6543
is very similar to that seen in archival data in 2001 October 
(Prog. Q108) and March 2006 (prog. F034). The spectrum in Fig. 1 also 
closely matches the
{\it ORPHEUS} spectrum obtained in 1993 September by Zweigle et al. 
(1997). 
Note for completeness that the $IUE$ high-resolution spectrum of
NGC~6543 is dominated by saturated P~Cygni profiles of
C{\sc iv}$\lambda\lambda$1548.20, 1550.77 and N{\sc v}
$\lambda\lambda$1238.82, 1242.80, plus well developed blueward
absorption in O{\sc v} $\lambda$1371.29 and the excited line of
N{\sc iv} $\lambda$1718.55 (e.g. Patriarchi and Perinotto 1997).

\begin{figure}
\centering
 {\scalebox{0.7}{\includegraphics{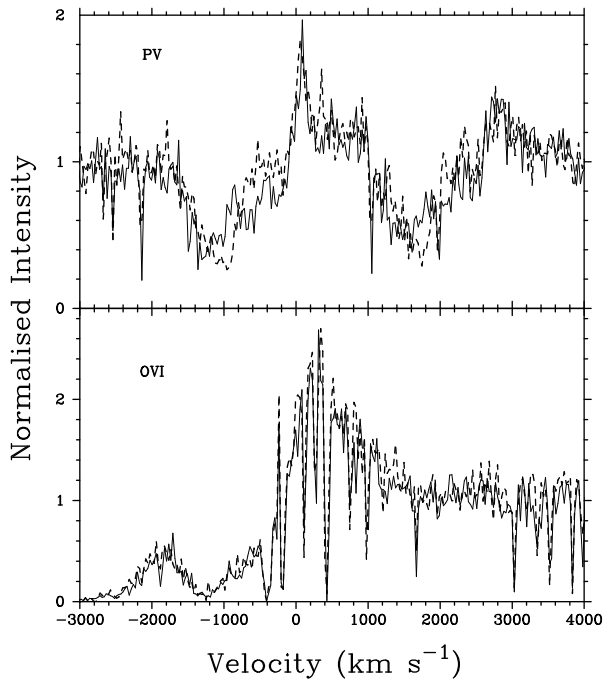}}}
 \caption{Fluctuations in the P{\sc v} resonance line profiles
compared to the more stable low-velocity
regions of O{\sc vi} $\lambda$1037.6}
\end{figure}


\section[]{Variability}

The P{\sc v} resonance doublet in NGC~6543 is a well developed
but unsaturated P~Cygni profile and thus offers the 
primary diagnostic of variable fast wind conditions in our dataset.
A pair of P{\sc v} line profiles separated by $\sim$ 2 days is shown in 
Fig. 2, to illustrate
what is essentially the maximum extent of profile fluctuations
evident.
A correction for a heliocentric systemic velocity of $-$66 km 
s$^{-1}$ (Schneider et al. 1983) was applied by subtracting
this value.
Mostly the flux changes are at $\sim$ 10 to 20{\%} of the
continuum level and occur over localised blueward velocity
regions, as opposed to the flux increasing or decreasing
simultaneously over the {\it entire} absorption 
trough. As is the case with massive OB stars, the
P{\sc v} P~Cygni {\it emission} profiles in NGC~6543 are
quite constant. The overall 
variance in the full 2007 time-series is certainly significant
between $\sim$ $-$500 to $-$1300 km s$^{-1}$.
The equivalent width of the P{\sc v} line is not conserved,
and fluctuates between $\sim$ 0.7 to 2.2 {\AA} (s.d. 0.3 {\AA})
for the blue component of the doublet (e.g. measured between
1111 to 1120 {\AA}).

Any changes in
the other spectral lines are rather subtle if present at all.
The lower panel in Fig. 2 shows the corresponding line profiles of
O{\sc vi}
$\lambda$1037.62. Even the low velocity (redward of 
$\sim$ $-$600  km s$^{-1}$) region of this line is likely
optically thick and
variance analysis does not support any changes in O{\sc vi}
that are significant above a 95{\%} confidence level.

\begin{figure*}
\centering
 {\scalebox{0.7}{\includegraphics{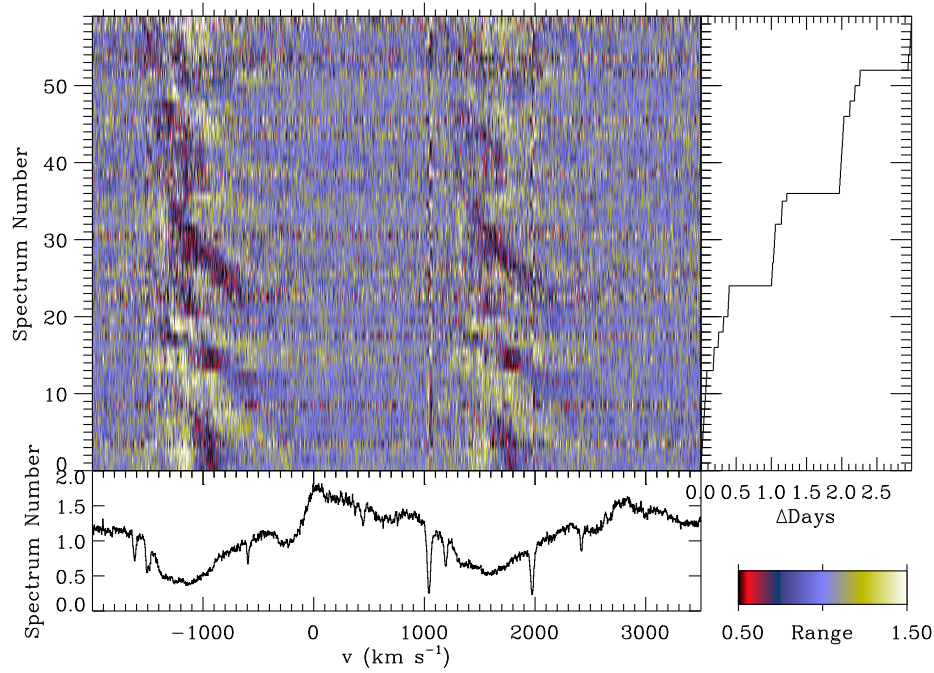}}}

\centering
 {\scalebox{0.75}{\includegraphics{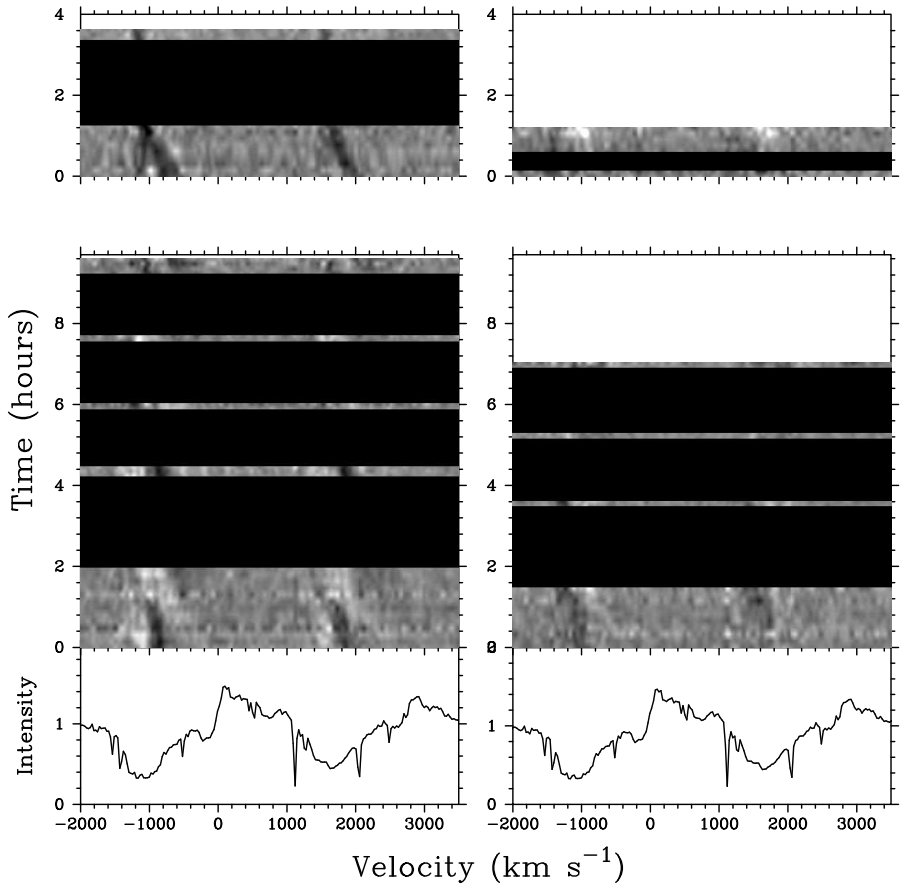}}}
 \caption{Dynamic spectrum representations of {\it 
organised}, hourly 
variability in P{\sc v} $\lambda\lambda$1118,1128. 
In the upper plot the ordinate is the sequential spectrum
number, and the panel on its right is a temporal plot showing
the relative time of each exposure. 
The grayscales in the lower plot show the {\it organised},
hourly variability. Starting from 
the
bottom, data in the pair of (lower) left-hand and right-hand panels span a
total of $\sim$ 1.22 days and 1.02 days, respectively. To enhance the
contrast in these images the individual spectra have been normalised
by the mean profile for the time-series.}
\end{figure*}


The data for P{\sc v} are by contrast very revealing. Representations
of the variability evident are displayed in Figure 3 as
dynamic spectra (i.e. two-dimensional images of individual
spectra ordered in time).
The images reveal clear evidence for
systematic and organised line profiles changes.
Specifically, localised (in velocity) absorption enhancements are seen
(darker shades in Fig. 3) migrating blueward from
 $\sim$ $-$400 to $\sim$ $-$1000 km s$^{-1}$.
We identify at least 4 separate episodes of recurring features.
The patterns in Fig. 3 are essentially identical to the
`discrete absorption components' (DACs) commonly seen in
UV resonance lines formed in the radiation-pressure-driven winds
of massive OB stars (e.g. Kaper et al. 1996; Prinja, Massa {\&}
Fullerton, 2002).
The lower left-hand panels in Fig. 3 focus on three of the
more prominent DAC-like features in NGC~6543, with
central velocities that increase
between $\sim$ $-$830 to $-$1010 km s$^{-1}$ over 1.1 hours,
$\sim$ $-$510 to $-$1290 km s$^{-1}$ over 6.0 hours, and
$\sim$ $-$700 to $-$1070 km s$^{-1}$ over 1.3 hours.

The DAC velocities above provide (linear)
accelerations of between 3 $\times$ 10$^{-2}$ km s$^{-2}$
to 8 $\times$ 10$^{-2}$ km s$^{-2}$. These values for NGC~6543
are up to a factor of 10 faster than
typical acceleration rates measured for features in OB star wind
(e.g. Prinja 1998). However the characteristic
radial flow time of the wind (which scales as $\sim$ 
R$_\star$/$v_\infty$) in NGC~6543 is only $\sim$ 5 minutes (Table 1),
compared to $\sim$ hours for O stars. The time-scales
associated with the DACs in NGC~6543 are therefore
significantly greater than the wind flushing time over the
line formation region, which suggests that it is very unlikely
the wind structures in the PN central star are due to processes entirely
intrinsic to the fast wind.
The DACs cannot, for example, be due to mass-conserved shells or blobs 
`riding' with the outflow from the star.
As is widely believed for OB stars, the perturbations more likely
arise as a result of inhomogeneities close to the stellar surface, such 
as pulsations or magnetic fields.
The empirical evidence provided here suggests that the physical
mechanism for initiating wind structure in the fast wind of
NGC~6543 may the same as that operating in massive, luminous
stars (see Sect. 6).

\section[]{Modulated or cyclic behaviour}

\begin{figure}
\centering
 {\scalebox{0.57}{\includegraphics{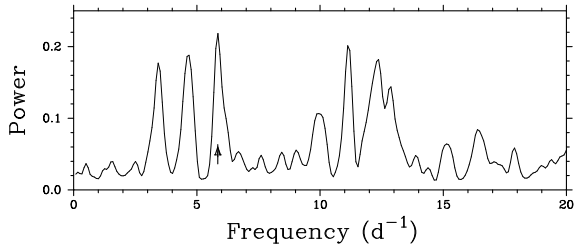}}}
 \caption{Power spectrum (in arbitrary units) for the
P{\sc v} $FUSE$ time-series. The arrow marks the
primary peak corresponding to
a period of 0.17 days in P{\sc v} (see Sect. 4).}
\end{figure}


In a few case studies based on very extended
$IUE$ spectroscopy it has been demonstrated that the DACs 
in OB stars are modulated or quasi-cyclic, on variability time-scales that 
may relate to stellar rotation periods (e.g. Massa et al. 1995, 
Fullerton et al. 1997, de Jong et al. 2001, Prinja et al, 2002).
From the viewpoint of nebular shapes, central star binarity,
and wind structure in line-driven winds,
it is obviously interesting to consider whether the optical
depth changes seen in the P{\sc v} lines of NGC~6543 are systematically
variable.
The rotation period of NGC~6543 is not known, and determinations of
the projected
rotation velocities are sparse for PN central stars in general.
Note also that the central star in our study is a member of
the ZZ Lep class of variable stars (e.g. Handler 2003).
Bell, Pollacco
{\&} Hilditch (1994) document differences of $\sim$ 0.01 mag in
5300 {\AA} CCD photometry though with little evidence for
periodic variations that are consistent over the four nights of
their observations.
Individual nights provided evidence for modulations over $\sim$2.6
hours and 3.2 hours.
Analysing the same dataset, Handler (2003)
cites a photometric modulation of 3.5 $\pm$ 1.5 hours.
The origin of the photometric changes in
NGC~6543 may relate to photospheric (pulsational) changes or
indeed the inner regions of the fast wind. The strong FUV
photospheric lines in our $FUSE$ data are relatively
fixed in velocity (to within $\pm$ 10 km s$^{-1}$)
and do not support any significant role for binary 
motion.

\begin{figure}
\centering
 {\scalebox{0.79}{\includegraphics{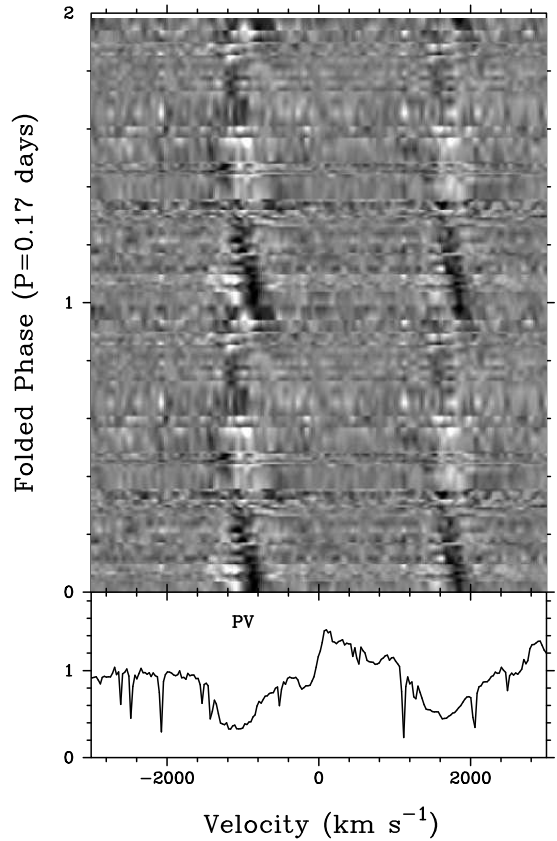}}}
 \caption{Individual P{\sc v} spectra are shown phased
over two cycles on a period of 0.17 days;
a clearly coherent behaviour is evident.}
\end{figure}


\begin{figure}
\centering
 {\scalebox{0.57}{\includegraphics{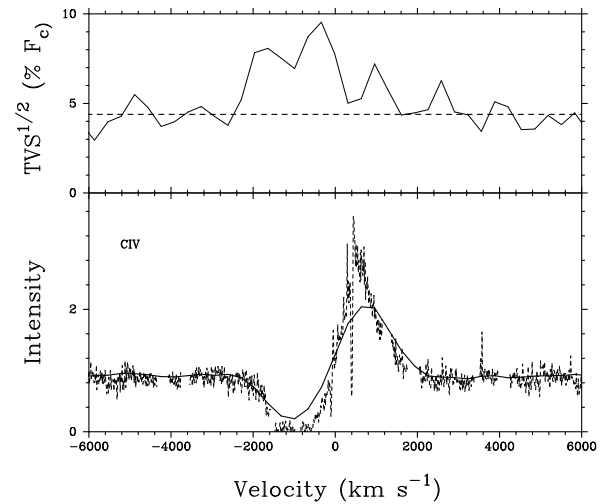}}}
 \caption{Variance spectrum for low-resolution $IUE$ time-series
of the C{\sc iv} $\lambda\lambda$1548,1550 line in NGC~6543.
The horizontal dashed line represents the 95{\%} confidence limit.
The lower panel shows the low-resolution line profile (solid
line) together with the a representative high-resolution
$IUE$ spectrum (dashed line).}
\end{figure}


 \begin{figure*}
 \centering
 {\scalebox{0.79}{\includegraphics{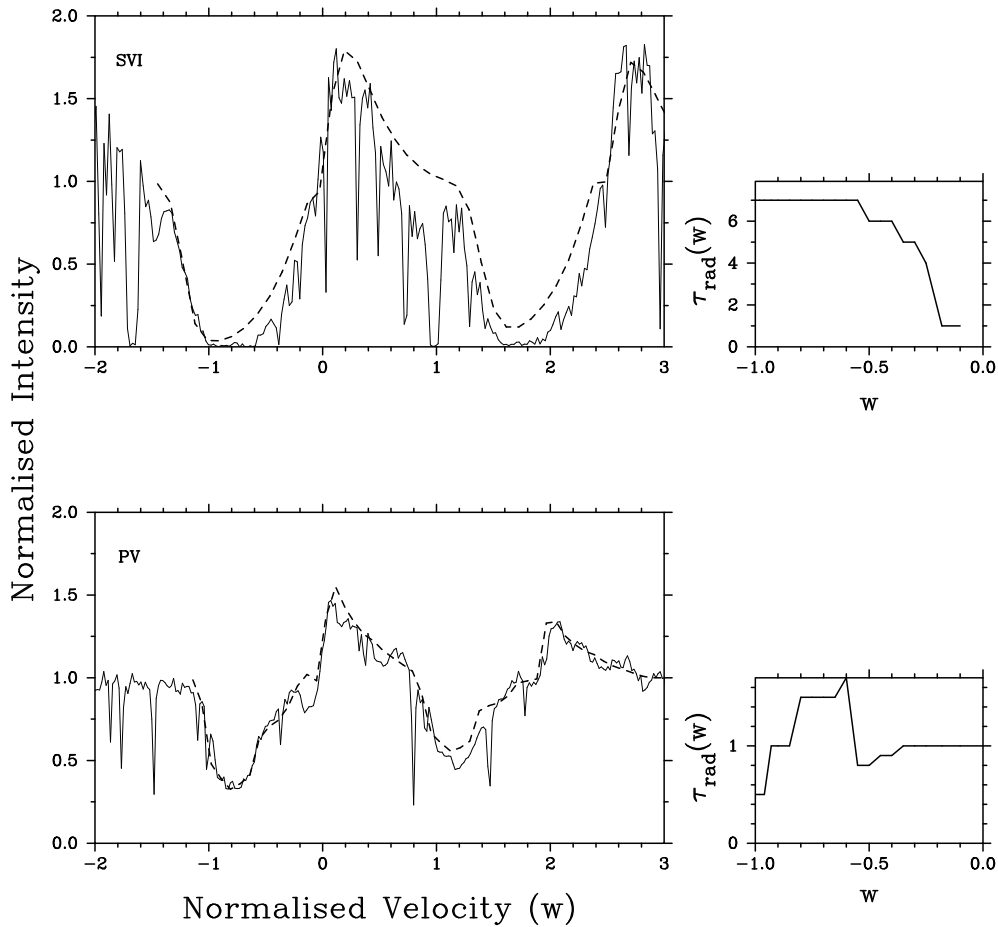}}}
 \caption{SEI model fits to the mean S{\sc vi} $\lambda\lambda$933, 945
 and P{\sc v} $\lambda\lambda$1118,1128. The adopted optical depth
bins as a function of normalised velocity (= $v/v_\infty$) are
shown in the right-hand panels.}
\end{figure*}


A periodogram analysis was carried out to search for evidence of 
repetitive
or cyclic properties in the P{\sc v} line profile changes.
The basic Fourier method involved uses the iterative CLEAN
algorithm (Roberts, Leh{\'a}r {\&} Dreher 1987) to 
deconvolve the features of the
window function from the discrete Fourier transform. 
(A gain of 0.5 with 100 iterations was used.) The power spectrum
sampled over $\sim$ $-$600 to $-$1100 km s$^{-1}$ (i.e. the range
of well-developed DACs) in P{\sc v} is shown in Fig. 4.
The data sampling is generally not intensive enough for a very robust
period search and the power in P{\sc v} is spread across several
frequencies. Nevertheless we identify as potentially interesting
the main power peak in  P{\sc v} at $\sim$ 5.85 days$^{-1}$,
corresponding to a period of $\sim$ 0.17 day. 
The half-width at half-maximum of the zeroth order peak in the
window function provides an approximate (and conservative) estimate
of the error in the frequency of $\sim$ 0.1 day$^{-1}$.
Independent determinations of the period sampled over 300
km s$^{-1}$ bins in the region between $-$600 to $-$1100 km s$^{-1}$
suggest an error in the frequency of $\sim$ 0.5 day$^{-1}$.
The adjacent peaks in the power spectrum (Fig. 4) are comparatively
less stable in similar velocity bin sampling.

Grayscales of individual spectra normalised to the mean are shown in
Fig. 5 phased on the 0.17 day period.
The diagnostic P{\sc v} data do clearly
show some coherency on the 0.17 day time-scale. The modulation
is represented by the occurrence of two sequential episodes
of migrating structures in the wind. (Recall that the total
length of the $FUSE$ time-series combined here is $\sim$ 3 days,
and therefore spans several `cycles'.)

Tentative evidence of variability on hourly time-scales can also been
gleaned by examining {\it low-resolution} (R $\sim$ 300)
$IUE$ spectra of the central star. We have retrieved
from the ESA $IUE$ archive a sequence of 38 spectra
(SWP54865 to SWP54960), which extend over 4.2 days. Despite
being saturated, the blue-wing of C{\sc iv} is apparently
highly variable (e.g. Patriarchi {\&} Perinotto, 1997). Furthermore,
studies of fluctuations in the extended blue-wings of saturated
P~Cygni profiles {\it have} (perhaps surprisingly) revealed
evidence for modulated behaviour (see e.g.
Howarth, Prinja {\&} Massa, 1995; Prinja et al., 1998).
We used high-resolution $IUE$ spectra of NGC~6543 that were
degraded (by convolution with a 5{\AA} full-width at half-maximum
Gaussian), to improve the internal consistency of the low-resolution
data and to normalise the continuum fluxes near C{\sc iv}.
The variance characteristics of the low-resolution $IUE$ C{\sc iv}
time-series is shown in Fig. 6. There is a double-peaked nature
to the TVS (e.g. Fullerton, Gies {\&} Bolton 1996) spectrum, and we elect 
to sample in particular the profile changes between $-$1000 and $-$2000 
km s$^{-1}$, since these are likely to represent variability in the
extreme blue-wings.
However, no signal
is detected in C{\sc iv} at the frequency identified for P{\sc v} 
modulations. The $IUE$ data quality prevent us from making any
firm conclusions as to whether any (quasi-)periodic
signal in NGC~6543 is long-lived.
We also attempted to
phase-fold the original {\it photometric} data of Bell et al. (1994,
obtained in 1993; Don Pollacco private communication) but did
not find any convincing evidence for a modulation on $\sim$ 0.17 days.

\section[]{Line-synthesis and optical depths}

In order to further quantify the physical properties of the
DACs in the wind of NGC~6543, the $FUSE$
resonance lines were modelled using the SEI method
(see Lamers, Cerrutti-Sola {\&} Perinotto 1987),
with the modified procedures of Massa et al. (2003).
The radial optical depth, $\tau_{\rm rad}(w)$, is treated as
21 independent velocity bins (each $\sim$ 0.05 $v_\infty$ wide)
to provide greater flexibility in matching absorption 
optical depths in a structured wind that is affected by variable features
such as DACs. We assume a standard parameterised velocity law
of the form,
\begin{equation}
w=w_0+(1-w_0)(1-1/x)^\beta ,
\end{equation}
where $w=v/v_\infty $ and $x=r/R_\star$.
The effect of a photospheric spectrum in the SEI profile fitting
is approximated by a Gaussian of full-width at half-maximum
= 250 km s$^{-1}$; its effect is only to improve the match
at low velocities ($\le$ 0.2 $v_\infty$) and does not impact
seriously on the overall wind parameters derived in this section.
Model fits to the mean P{\sc v} profile in our time-series is
shown in Fig. 7, where a very good overall match is achieved
for $v_\infty$ = 1400 km s$^{-1}$, $\beta$ = 1, and
$v_{\rm turb}$ (the small-scale velocity dispersion parameter) = 0.05.
Note that the redward component of P{\sc v} may be affected by the
(neglected) contribution due to
Si{\sc iv} $\lambda\lambda$1122.5, 1128.3.
These values are also consistent for reproducing the blue-wing of 
a much stronger (saturated) profile such as S{\sc vi} (Fig. 7).

The central star is H-rich, with normal solar abundance ratios
though Carbon may be overabundant (e.g. Georgiev et al. 2006).
The SEI model of P{\sc v} in Fig. 7 therefore gives
a (mean) value for the product of mass-loss rate and ionization
fraction, $\mdot\,q(P^{4+})$, integrated over
0.2 to 0.9 $v_\infty$, of $\sim$ 5.0 $\times$ 
10$^{-9}$ M$_\odot$ yr$^{-1}$ (for parameters in Table 1).
Over the velocity range where the peak variance occurs in
P{\sc v} (i.e. $\sim$ 0.5 to 0.9 $v_\infty$), SEI fits to the
cases of minimum and maximum observed absorption
yield $\mdot\,q(P^{4+})$ values of
$\sim$ $\sim$ 1.9 $\times$ 10$^{-9}$ M$_\odot$ yr$^{-1}$
and 3.5 $\times$ 10$^{-9}$ M$_\odot$ yr$^{-1}$.

\begin{figure}
\centering
 {\scalebox{0.57}{\includegraphics{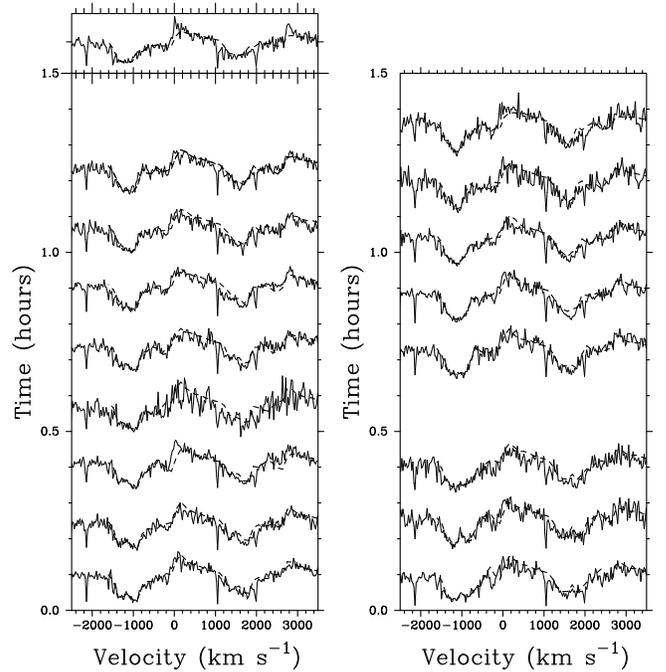}}}
 \caption{SEI model fits (dashed lines) to sequences of
P{\sc v} line profiles where two of the prominent DACs
are migrating between $\sim$ $-$600 km s$^{-1}$ to $-$1100 km s$^{-1}$;
i.e. the lower left-hand panels in Fig. 3. The isolated profile fitted
in the upper left panel is a representative `no-DAC' (maximum flux)
case.}
\end{figure}

\begin{figure}
\centering
 {\scalebox{0.57}{\includegraphics{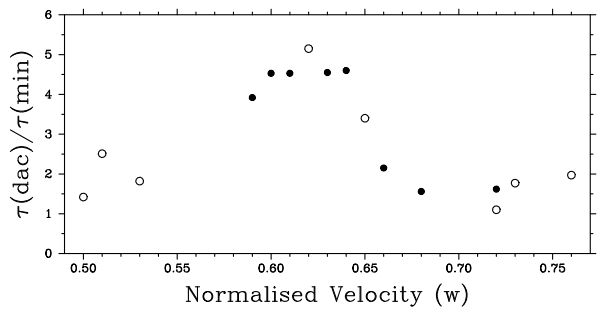}}}
 \caption{Ratio of optical depth in the DAC to minimum optical
depth at that velocity from SEI fits, for two independent episodes
(i.e. open and closed symbols) identified in Figs. 3 and 8.}
\end{figure}


The fact that the SEI model can match well the
overall morphology of the
observed absorption in P{\sc v} permits us to model individual
line profiles in the $FUSE$ time-series. Specifically we focus
on the data covering the two most prominent DAC-like features
in lower left-hand panels of Fig. 3 (in both cases the features
start from the normalised time of 0.0 hours; the corresponding
FUSE exposures are F03401050012 to F03401050082
and F03401060042 to F03401060122). The fits to these
spectra, which are varying due to the occurrence of DACs, can then
be compared with a `no-DAC' case i.e. a spectrum
representative of maximum flux over most velocities
(i.e. F03401080072).

\begin{figure*}
\centering
 {\scalebox{0.7}{\includegraphics{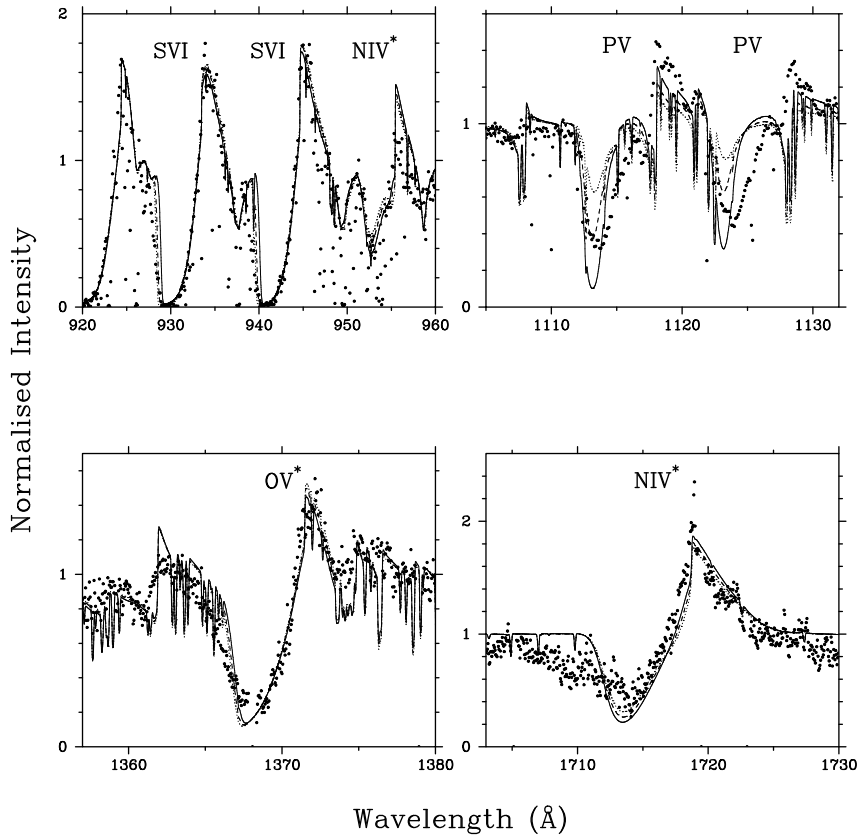}}}
 \caption{Key diagnostics lines in the $FUSE$ and high-resolution
$IUE$ spectra (dots) of the central star are shown with CMFGEN model fits
for $f_\infty$ = 0.06 (solid line), 0.08 (dashed line) and
0.1 (dotted line). Notice the particular sensitivity of
P{\sc v} to the clumping factor.}
\end{figure*}


The SEI models fits to the P{\sc v} (time-series) line profiles are shown 
in Figure 8.
The flexible optical depth bins are designed to allow matches
to a wind profile that is obviously not `smooth', but is instead
distorted due to density, ionization or velocity features
(see e.g. Massa et al. 2003). The good overall reproduction in
Fig. 8 permits us to extract optical depths at the specific velocity
of the migrating DACs, and normalise them to the no-DAC case.
Ratios of $\tau$(DAC)/$\tau$(min) (obtained from fits to
the individual profiles) are plotted as 
function
of normalised velocity ($v$/$v_\infty$) in Fig. 9.
The plot therefore shows the optical depth evolution of
two independent DAC episodes as a function of their central
velocities.
In both cases the optical depth ratio due to DACs rises to a maximum
of $\sim$ 5 at $\sim$ 0.62 $v_\infty$, and drops to $\sim$ 2
toward 0.75 $v_\infty$. The indication that the maximum contrast
of the DACs is reached at intermediate velocities is somewhat
unexpected.
The increase in Sobolev optical depth at these positions may be due to a 
plateau in the wind velocity law, i.e. a radially extended region of
near-constant velocity. Wind material flows through the plateau,
and the observed accelerations of DACs are then associated with
the evolution (in velocity and radial extent) of the plateau as
a function of time.
The origin of such a plateau could, for example,
relate to the formation of co-rotating interactions regions
in the outflow (see e.g. Cranmer {\&} Owocki, 1996, and discussion
in Sect. 6).

To gain some insights into the ionization balance and wind clumping
in NGC~6543, we also employed the unified model atmosphere
code CMFGEN (Hiller {\&} Miller, 1998), which calculates non-LTE
models of spherically symmetric winds with multi-level atoms.
The fundamental parameters listed in Table 1 were adopted, together
with log $g$ = 4.2 and a He/H abundance ratio (by number) of 0.1 (e.g. 
Georgiev
et al. 2006). A parametric treatment of wind clumping is implemented
in CMFGEN, in terms of the volume filling factor, $f$, defined
as
\begin{equation}
f = f_\infty + (1 - f_\infty)exp(-v/v_{cl}),
\end{equation}
where $v_{cl}$ is the velocity at which clumping starts, and which we
adopt as 35 km s$^{-1}$ (i.e. above the sonic point).
Our objective here is primarily
to investigate the spectral signatures of clumping in CMFGEN models
applied to the fast wind of NGC6543; we do not attempt a detailed
exercise in fitting several lines.


\begin{table*}
 \centering
\caption{Comparison of NGC~6543 and mid-O star DAC properties.}
  \begin{tabular}{lll}
  \hline
DAC measure & NGC~6543 & O stars (e.g. O7 III)  \\
\hline

$v_{\rm initial}$ & $\sim$ 0.3 $v_\infty$ & $\le$ 0.3 $v_\infty$ \\
Recurrence & yes (over $\sim$ hours) & yes (over $\sim$ days) \\
(dv/dt)$\times$(R$_\star$/$v_\infty$) & $\sim$ 12$-$24 km s$^{-1}$& 
$\sim$ 10 km s$^{-1}$ \\
maximum $\tau$(DAC)/$\tau$(min) & $\sim$ 5 & $\sim$ 5 \\
Quasi-periodic & yes; P $\sim$ 0.14 days & yes; e.g. P $\sim$ days \\
FWHM/$v_\infty$ (initial) & $\sim$ 0.15 $v_\infty$ & $\sim$ 0.3 
$v_\infty$ \\
FWHM/$v_\infty$ (final) & $\le$ 0.05 $v_\infty$ & $\le$ 0.1
$v_\infty$ \\
\hline

\end{tabular}
\end{table*}


Models were run for cases of the clump volume filling factor,
$f_\infty$ = 0.06, 0.08 and 0.1, and the synthesised spectra are
compared in Fig. 10 to key wind lines in the $FUSE$ and high-resolution 
$IUE$ data of the central star. For a mass-loss rate of
6 $\times$ 10$^{-8}$ M$_\odot$ yr$^{-1}$ the models provide good
overall fits to the observed resonance and excited state line profiles.
It is interesting to note that the P{\sc v} lines are particularly
sensitive to the clumping factor, more so than O{\sc v} and
N{\sc iv}. Our data favour the case of $f_\infty$ = 0.08, though
of course we neglect here the consequences of these parameters
for reproducing diagnostic optical lines in the central star of
NGC~6543, such as He{\sc ii} $\lambda$4686 and C{\sc iv} $\lambda$5801. 
Figure 11 displays
the predicted ionization fractions of P$^{3+}$, P$^{4+}$, and
P$^{5+}$ as a function of velocity, and for the three different
clumping factors. The dominance of P$^{5+}$ at low velocities results
in the mis-match with the observed profile seen in this region (Fig. 10).
The ion fractions of P$^{3+}$ and P$^{4+}$ increase
over all velocities for lower values of $f_\infty$,
with an increased recombination from P$^{5+}$ to P$^{4+}$
particularly evident at the highest velocities.
The CMFGEN models
run here generally predict C$^{4+}$ and P$^{5+}$ as dominant
(i.e. $q_i$ $>$ 0.97 over $\sim$ 0.2 to 0.9 $v_\infty$), and
S$^{5+}$ and O$^{4+}$ close to dominant (i.e. $q_i$ $\sim$ 0.6
and 0.2, respectively).

Our results indicate that P{\sc v} is a sensitive spectral diagnostic
of {\it small-scale} clumping in the fast wind, together with the
large-scale structures represented by the DACs discussed in
Sect. 3. We highlight the need to adopt fairly small volume filling
factors in order to match the P{\sc v} resonance line doublet.
P$^{4+}$ is a trace ion, just below dominant (Fig. 11), and is
therefore extremely sensitive to changes as parameterised here, since
its abundance depends strongly on density. This is in contrast to
the results for O stars presented by Bouret et al. (2005), where
P$^{4+}$ is dominant and not as sensitive to small fluctuations
in density.

\section[]{Discussion}

The empirical evidence presented here suggests that (i) the fast wind
of NGC~6543 is highly variable on time-scales of hours, (ii) the
wind changes are systematic in terms of recurrent absorptive optical
depth structures that migrate blueward, and (iii) the temporal
behaviour of the P{\sc v} resonance line P~Cygni profiles is
very similar to that due to the presence of `discrete absorption
components' (DACs) in luminous OB stars.
A comparison between the DAC properties of NGC~6543 and a
`representative' mid-O giant is presented in Table~3, where the latter
is summarised from various studies (see e.g. those cited in Sect. 3).
We note that the fundamental characteristics such as initial
detection at low Doppler velocities, recurrence, modulation, and
optical depth contrast are broadly similar for both cases; the same
is also true for the (linear) acceleration scaled by the flow
time-scale (R$_\star$/$v_\infty$). There is some indication that the
line-of-sight velocity dispersion of the DACs in NGC~6543 is somewhat
smaller than for a typical O-type giant (such as $\xi$ Per or
68 Cygni).

\begin{figure}
\centering
 {\scalebox{0.57}{\includegraphics{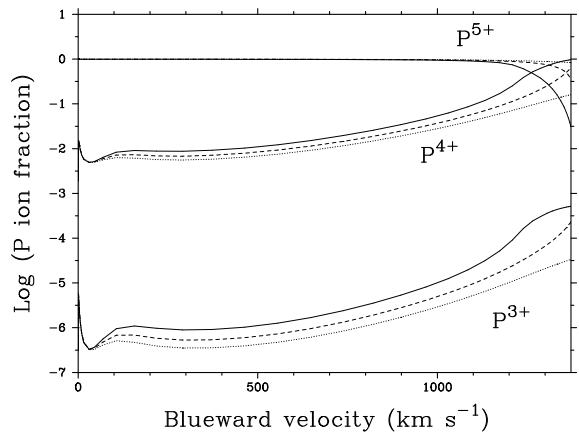}}}
 \caption{Log of the P{\sc iv}, P{\sc v} and
P{\sc vi} ion fractions as a function 
of blueward velocity, for cases of $f_\infty$ = 0.06 (solid line),
0.08 (dashed line) and 0.1 (dotted line).}
\end{figure}


Our interpretation of these results therefore is that the physical
mechanism responsible for the DAC structures in the fast wind
of NGC~6543
is likely the same as that acting in massive OB stars.
There remain several uncertainties in our understanding of this
phenomena in massive stars, but in the particular case where
quasi-periodic wind variability is detected, several authors have
promoted a scenario in terms of co-rotating interaction regions
(CIRs; e.g. Cranmer {\&} Owocki 1996; Fullerton et al. 1997;
de Jong et al. 2001). The notion is that the DACs are spectroscopic
signatures of spatial structures in the wind that are causally connected
to stellar surface irregularities. These photospheric inhomogeneities
cause the wind from different longitudinal sectors on the stellar
surface to emerge with different densities and/or velocities.
The consequence is to form different adjacent streams that meet
to create spiral-shaped CIRs.
Key observational properties of the DACs can be matched with this
model, with variable optical depth enhancements arising from the
combination of a plateau in the radial velocity as well as a
density perturbation.
In this model, the variations in
absorption which occurs over significantly larger time-scales than
the wind flow speed are due to gas flowing through a semi-permanent,
co-rotating structure. The rotation rate of the central star
in NGC~6543 is unfortunately not known, and we
are unable to comment further 
in this context on the significance of the $\sim$ 0.17-day period
identified in Sect. 4. 
In the case of OB stars the recurrence
of DACs is mostly only quasi-cyclic, and has only rarely been
demonstrated to be an integer relation to the stellar rotation
rate.
Nevertheless, our results provide the constraint that the mechanism
for forming coherent perturbations in the outflows is apparently 
operating equally in the radiation-pressure-driven winds of widely
differing momenta ($\mdot$$v_\infty$$R_\star^{0.5}$) and
flow times, as represented by OB stars and CSPN.

As regards the `trigger' provided by photospheric irregularities
for the growth of wind structure,
the action of surface velocity fields due to pulsation and/or
ordered magnetic fields may be just as applicable to CSPN as it
is to OB stars. Multi-period pulsation modes are thought to be the
cause of short-period ($\sim$ hourly) variability seen in the
photometric measurements of several central stars, including
both H-rich and H-deficient cases (e.g. Handler et al. 1997;
Ciardullo {\&} Bond 1996). Kilogauss strength magnetic fields
have also been detected (using spectropolarimetry) that are
associated with the PN central stars (e.g. Jordan, Werner {\&}
O'Toole 2005), where they may be located mostly in the
envelope, as opposed to the degenerate core, and can be affected by
mass-loss. One caveat here is that the link between pulsation or
magnetic fields and repetitive wind structures in OB stars has
thus far proved observationally very challenging and generally
unconstrained; establishing such a connection in CSPN is going to be
even more difficult!

The {\it ab inito} CMFGEN calculations presented in Sect. 5 favour
a small clump volume filling factor, f$_\infty$ $\sim$ 0.08. Whether
or not the parameterisation of clumping incorporated in CMFGEN is
physically meaningful, we note that the high sensitivity of the
P{\sc v} lines to the clumping factor is again similar to that
apparent for the same line in O-type stars. Following the clumping
studies of e.g. Markova et al. (2004), Bouret et al. (2005), 
Fullerton et al. (2006), our results for NGC~6543 would imply
a reduction by a factor $\sim$ 4 (i.e. 1/$\sqrt{f_\infty}$) 
in the mass-loss rate due to the clumped fast wind compared to 
smooth-wind
models. The analysis of {\it optical} lines with the non-LTE
code FASTWIND by Kudrtizki, Urbaneja {\&} Puls 2006) also points
to substantially clumped CSPN fast winds.
Downward revisions in the central star mass-loss rate would affect
the dynamics of the hot central cavity excavated by the fast wind,
thus modifying its interaction with the surrounding nebula,
including contributions to the X-ray emission.
Naturally our far-UV
analyses needs to be extended to cover a large sample of
central stars to permit firm assessments of any revisions to CSPN
mass-loss rates, and most importantly, tested for consistency between UV 
and optical diagnostics of clumping.
Finally, looking ahead, the vastly increased sensitivity soon to
be offered at radio wave-bands by the upgraded e-MERLIN and
EVLA facilities raises new possibilities for measuring the
free-free emission from the fast wind itself and thus provide
a powerful new constraint on the central star mass-loss.

\section*{Acknowledgments}
Based on observations made with the NASA-CNES-CSA Far Ultraviolet
Spectroscopic Explorer. FUSE is
operated for NASA by the Johns Hopkins University under NASA contract
NAS5-32985.
DLM acknowledges support through NASA contract NNG07EF12P to SGT, Inc.
SEH is grateful for studentship support from STFC (formerly PPARC).

\bsp

\label{lastpage}

\end{document}